\documentclass[preprint,aps]{revtex4}

\usepackage{graphicx}
\usepackage{dcolumn}
\usepackage{bm}

\begin{document}

\title{Spontaneous CP Violating Phase as The CKM Matrix Phase}

\author{Shao-Long Chen$^1$, N. G. Deshpande$^2$, Xiao-Gang He$^1$, Jing Jiang$^2$ and
Lu-Hsing Tsai$^1$}
\affiliation{$^1$Department of Physics and Center for Theoretical Sciences, National Taiwan University, Taipei, Taiwan\\
$^2$Institute of Theoretical Science, University of Oregon, Eugene, OR 97403, USA}

\date{\today} 

\begin{abstract}
We propose that the CP violating phase in the CKM mixing matrix is
identical to the CP phases responsible for the spontaneous CP
violation in the Higgs potential. A specific multi-Higgs
model with Peccei-Quinn (PQ) symmetry is constructed to realize
this idea. The CP violating phase does not vanish when all Higgs
masses become large. There are flavor changing neutral current
(FCNC) interactions mediated by neutral Higgs bosons at the tree
level. However, unlike general multi-Higgs models, the FCNC Yukawa
couplings are fixed in terms of the quark masses and CKM mixing
angles. Implications for meson-anti-meson mixing, including recent
data on $D-\bar D$ mixing, and neutron electric dipole moment
(EDM) are studied. We find that the neutral Higgs boson masses can
be at the order of one hundred GeV. The neutron EDM can be close
to the present experimental upper bound.
\end{abstract}

\maketitle

\newpage

\section{Introduction}

The origin of CP violation is one of the outstanding problems of
modern particle physics. There have been several experimental
measurements of CP violation~\cite{pdg}. All of them are
consistent with the Cabbibo-Kobayashi-Maskawa (CKM)
model~\cite{cab,km}, where the source of CP violation comes from
the phase~\cite{km} $\delta_{KM}$ in the CKM mixing matrix for
quarks. A successful model of CP violation at the leading order
should have the successful features of the CKM model. It is
important to understand the origin of CP violation. An interesting
proposal due to T.-D. Lee was that CP is spontaneously
violated~\cite{tdlee}. The popular Weinberg model~\cite{weinberg1}
of spontaneous CP violation model has problems~\cite{problem,bigi}
with data and has been decisively ruled out by CP violating
measurement in B decays~\cite{pdg}. Spontaneous CP violation in
the Left-Right models has also been ruled out for the same
reason\cite{Ball:1999mb}. In this work we restore the idea that CP
is broken spontaneously and the phase $\delta_{KM}$ is the same as
the phase $\delta_{spon}$ that causes spontaneous CP violation in
the Higgs potential.  We construct specific models to realize this
idea. The main difference of our models lies in how the CP
violating phase in the CKM matrix is identified~\cite{other}.

Let us start by describing how a connection between $\delta_{KM}$
and $\delta_{spon}$ can be made. It is well known that to have
spontaneous CP violation, one needs two or more Higgs doublets
$\phi_i$. Consider the following Yukawa couplings with multi-Higgs
doublets,
\begin{eqnarray}
L_Y = \bar Q_L (\Gamma_{u1} \phi_1 + \Gamma_{u2} \phi_2) U_R +
\bar Q_L \Gamma_d \tilde \phi_d D_R + h.c.\;,
\end{eqnarray}
where $Q_L$, $U_R$ and $D_R$ are the left-handed doublet,
right-handed up and right-handed down quarks, respectively.
Generation indices are suppressed. $\tilde \phi_d = -i\sigma_2
\phi_d^*$ and $\phi_d$ may be one of the $\phi_{1,2}$ or another doublet Higgs
field. The Yukawa couplings $\Gamma_{u1,u2,d}$ must be real if CP is
only violated spontaneously.

The Higgs doublets when expressed in terms of the component fields
and their vacuum expectation values (vev) $v_i$ are given by
\begin{eqnarray}
\phi_i = e^{i\theta_i}H_i = e^{i\theta_i}\left ( \begin{array}{c}
{1\over \sqrt{2}}(v_i +R_i +i A_i)\\
h^-_i\end{array} \right ).
\end{eqnarray}
The quark mass terms in the Lagrangian are
\begin{eqnarray}
L_m &= &-\bar U_L \left [M_{u1}e^{i\theta_1} + M_{u2} e^{i\theta_2}\right ]U_R
- \bar D_L M_d e^{-i\theta_d}D_R + h.c.\;,
\end{eqnarray}
where $M_{ui} =-\Gamma_{ui} v_i / \sqrt{2}$.

The phases $\theta_1$ and $\theta_d$ can be absorbed by redefining
the fields $U_R$ and $D_R$. However, the phase difference $\delta
= \theta_2 - \theta_1$ cannot be removed and it depends on the
Higgs potential. A non-zero $\delta$ indicates spontaneous CP
violation, $\delta = \delta_{spon}$. Without loss of generality,
we work in the basis where $D_L$, $D_R$ are already in their mass
eigenstates. In this basis the down quark mass matrix $M_d$ is
diagonalized, which will be indicated by $\hat M_d$. In general
the up quark mass matrix $M_u = M_{u1} + e^{i\delta}M_{u2}$ is not
diagonal. Diagonalizing $M_u$ produces the CKM mixing matrix. One
can write $\hat M_u = V_{CKM} M_uV^\dagger_R$.  Here $V_{CKM}$ is
the CKM matrix and $V_R$ is an unknown unitary matrix.  A direct
identification of the phase $\delta_{spon}$ with the phase
$\delta_{KM}$ in the CKM matrix is not possible in general at this
level. There are, however, classes of mass matrices which allow
such a connection. A simple example is provided by setting $V_R$
to be the unit matrix. With this condition,  $M_u =
V^\dagger_{CKM} \hat M_u$. One then needs to show that
$V_{CKM}^\dagger$ can be written as
\begin{eqnarray}
V_{CKM}^\dagger = (M_{u1} + e^{i\delta} M_{u2})\hat M^{-1}_u.
\label{mvkm}
\end{eqnarray}
Expressing the CKM matrix in this form is very suggestive. If
$V_{CKM}$ (or $V_{CKM}^\dagger$) can always be written as a sum of two
terms with a relative phase, then the phase in the CKM matrix can be
identified with the phase $\delta$.

We now demonstrate that it is the case by using the Particle Data
Group (PDG) parametrization as an example.  To get as close as to
the form in Eq.~(\ref{mvkm}), we write the PDG CKM matrix
as~\cite{pdg}
\begin{eqnarray}
V_{CKM} = \left(
\begin{array}{lll}e^{-i\delta_{13}}&0&0\\0&1&0\\0&0&1\end{array}\right)
\left( \begin{array}{lll}
c_{12}c_{13}e^{i\delta_{13}}&s_{12}c_{13}e^{i\delta_{13}}&s_{13}\\
-s_{12}c_{23}-c_{12}s_{23}s_{13}e^{i\delta_{13}}&c_{12}c_{23}-s_{12}s_{23}s_{13}
e^{i\delta_{13}}&s_{23}c_{13}\\
s_{12}s_{23}-c_{12}c_{23}s_{13}e^{i\delta_{13}}&-c_{12}s_{23}-s_{12}c_{23}s_{13}
e^{i\delta_{13}}&c_{23}c_{13}
\end{array} \right ),
\end{eqnarray}
where $s_{ij} = \sin\theta_{ij}$ and $c_{ij}=\cos\theta_{ij}$.

Absorbing the left matrix into the definition of $U_L$ field, we
have
\begin{eqnarray}
M_{u1} = \left ( \begin{array}{ccc} 0& - s_{12}c_{23}&s_{12}s_{23}\\
0&c_{12}c_{23}&-c_{12}s_{23}\\
s_{13}&s_{23}c_{13}&c_{23}c_{13}\end{array} \right ) \hat
M_u\;,\;\; M_{u2} = \left ( \begin{array}{ccc}
c_{12}c_{13}&-c_{12}s_{23}s_{13}&-c_{12}c_{23}s_{13}\\
s_{12}c_{13}&-s_{12}s_{23}s_{13}&-s_{12}c_{23}s_{13}\\
0&0&0\end{array} \right ) \hat M_u\;,
\end{eqnarray}
and $\delta = -\delta_{13}$. We therefore find that it is possible
to identify the CKM phase with that resulting from spontaneous CP
violation. Note that as long as the phase $\delta$ is not zero, CP
violation will show up in the charged currents mediated by W
exchange. The effects do not disappear even when Higgs boson
masses are all set to be much higher than the $W$ scale.
Furthermore, $M_{1,2}$ are fixed in terms of the CKM matrix
elements and the quark masses, as opposed to being arbitrary in
general multi-Higgs models.

We comment that the solution is not unique even when $V_R$ is set
to be the unit matrix.  To see this, one can take another
parametrization for the CKM matrix, such as the original
Kobayashi-Maskawa (KM) matrix~\cite{km}. More physical requirements
are needed to uniquely determine the connection. The
phenomenological consequences will therefore be different. We will
come back to this when we look at phenomenology of models. The key
point we want to establish is that there are solutions where the
phase in the CKM matrix can be identified with the phase causing
spontaneous CP violation in the Higgs potential.

The mass matrices $M_{u1}$ and $M_{u2}$ can be written in a parametrization
independent way in terms of the eigen-mass matrix $\hat M_u$, the
CKM matrix, and the phase $\delta$,
\begin{eqnarray}
M_{u1}&=& V^\dagger_{CKM}\hat M_u - {e^{i\delta}\over \sin\delta} Im(V^\dagger_{CKM})\hat M_u\;,\nonumber\\
M_{u2}&=& {1\over \sin\delta} Im(V^\dagger_{CKM}) \hat M_u\;.
\end{eqnarray}

Alternatively, a model can be constructed with two Higgs doublets
couple to the down sector and one Higgs doublet couples to the up
sector to have,
\begin{eqnarray}
L_Y = \bar Q_L \Gamma_u \phi_uU_R + \bar Q_L( \Gamma_{d1} \tilde
\phi_1 +\Gamma_{d2} \tilde \phi_2) D_R + h.c.\;.
\end{eqnarray}
In this case $M_{di} = -\Gamma_{di} v_i/\sqrt{2}$, and
\begin{eqnarray}
M_{d1}&=& V_{CKM}\hat M_d + {e^{-i\delta}\over \sin\delta} Im(V_{CKM})\hat M_d\;,\nonumber\\
M_{d2}&=&- {1\over \sin\delta} Im(V_{CKM}) \hat M_d\;.
\end{eqnarray}

We denote the above two possibilities as Model a) with two Higgs
doublets coupled to the up sector, and Model b) with two Higgs
doublets coupled to the down sector.

\section{Model Building}

A common problem for models with spontaneous CP violation is that
a strong QCD $\theta$ term will be generated~\cite{bigi}.
Constraint from neutron dipole moment measurement will rule out
spontaneous CP violation as the sole source if there is no
mechanism to make sure that the $\theta$ term is small enough if
not zero. The models mentioned above face the same problem. We
therefore supplement the model with a Peccei-Quinn (PQ)
symmetry~\cite{pq} to ensure a small $\theta$.

To have spontaneous CP violation and also PQ symmetry
simultaneously, more than two Higgs doublets are
needed~\cite{hvgn}. For our purpose we find that in order to have
spontaneous CP violation with PQ symmetry at least three Higgs
doublets $\phi_i = e^{i\theta_i}H_i$ and one complex Higgs singlet
$\tilde S = e^{i\theta_s}S = e^{i\theta_s}(v_s + R_s + i
A_s)/\sqrt{2}$ are required. The Higgs singlet with a large vacuum
expectation value (vev) renders the axion from PQ symmetry
breaking to be invisible~\cite{invisible,kk}, thus satisfying
experimental constraints on axion couplings to fermions. We will
henceforth work with models with an invisible
axion~\cite{invisible}.

The PQ charges for Models a) and b) are as follows
\begin{eqnarray}
\mbox{Model a)}&&Q_L : 0\;,\;\;U_R: -1\;,\;\;D_R:
-1\;,\;\;\phi_{1,2}:
+1\;,\;\;\phi_d=\phi_3: -1;\nonumber\\
\mbox{Model b)}&&Q_L : 0\;,\;\;U_R: +1\;,\;\;D_R:
+1\;,\;\;\phi_{1,2}: +1\;,\;\;\phi_u=\phi_3: -1.
\end{eqnarray}
In both cases, $\tilde S$ has PQ charge $+2$. For leptons, the PQ
charges can have different assignments. For example: $L_L:
0\;,\;\;e_R: -1$ or $L_L: 0\;,\;\;e_R: +1$.

For both models a) and b), the Higgs potentials have the same form
which is given by
\begin{eqnarray}
V&=& -m^2_1 H_1^\dagger H_1-m^2_2 H_2^\dagger H_2-m^2_3
H_3^\dagger H_3-m^2_{12} (H_1^\dagger H_2e^{i(\theta_2 -
\theta_1)}+ h.c.)-m^2_s S^\dagger S\nonumber\\
&+&\lambda_1 (H_1^\dagger H_1)^2+\lambda_2 (H_2^\dagger H_2)^2
+\lambda_t (H_3^\dagger H_3)^2 + \lambda_s (S^\dagger
S)^2\nonumber\\
&+&\lambda_3 (H_1^\dagger H_1)(H^\dagger_2H_2)+ \lambda'_3
(H_1^\dagger H_1)(H^\dagger_3H_3)+\lambda''_3 (H_2^\dagger
H_2)(H^\dagger_3H_3)\nonumber\\
&+&\lambda_4 (H_1^\dagger H_2)(H^\dagger_2H_1)+ \lambda'_4
(H_1^\dagger H_3)(H^\dagger_3H_1)+\lambda''_4 (H_2^\dagger
H_3)(H^\dagger_3H_2)\nonumber\\
&+&{1\over 2}\lambda_5
((H^\dagger_1H_2)^2e^{i2(\theta_2-\theta_1)}+h.c.) +\lambda_6
(H^\dagger_1 H_1)(H^\dagger_1 H_2 e^{i(\theta_2-\theta_1)}+h.c.)
\nonumber\\
&+&\lambda_7 (H^\dagger_2 H_2)(H^\dagger_1 H_2
e^{i(\theta_2-\theta_1)}+h.c.)+\lambda_8 (H^\dagger_3
H_3)(H^\dagger_1 H_2 e^{i(\theta_2-\theta_1)}+h.c.)\nonumber\\
&+& f_1 H^\dagger_1 H_1 S^\dagger S + f_2 H^\dagger_2 H_2
S^\dagger
S + f_3 H^\dagger_3 H_3 S^\dagger S + d_{12} (H^\dagger_1 H_2e^{i(\theta_2-\theta_1)}
+ H^\dagger_2 H_1e^{-i(\theta_2 -\theta_1)}) S^\dagger S\nonumber\\
&+& f_{13} (H^\dagger_1 H_3 S e^{i(\theta_3 + \theta_s -
\theta_1)} + h.c.)+f_{23} (H^\dagger_2 H_3 S e^{i(\theta_3 +
\theta_s - \theta_2)} + h.c.)\;.
\end{eqnarray}
Only two phases occur in the above expression, which we choose to be
$\delta =
\theta_2-\theta_1$ and $\delta_s = \theta_3 +\theta_s - \theta_2$.
The phase $\theta_3+\theta_s - \theta_1$ can be written as $\delta
+ \delta_s$. Differentiating with respect to $\delta_s$ to get one of the
conditions for minimization of the potential, we get
\begin{eqnarray}
&&f_{13}v_1v_3 v_s \sin(\delta_s + \delta) + f_{23} v_2v_3 v_s
\sin\delta_s = 0\;.
\end{eqnarray}
We see that $\delta$ and $\delta_s$ are related with
\begin{eqnarray}
\tan\delta_s = -{f_{13} v_1 \sin\delta\over f_{23}v_2 + f_{13} v_1
\cos\delta}\;.
\end{eqnarray}

Therefore, $\delta$ is the only independent phase in the Higgs
potential. A non-zero $\sin\delta$ is the source of spontaneous CP
violation and also the only source of CP violation in the model.

In this model the Goldstone fields $h_w$ and $h_z$ that are ``eaten'' by $W$
and $Z$, and the axion field are given by
\begin{eqnarray}
&&h_w = {1\over v}(v_1 h^-_1 + v_2 h^-_2 + v_3
h^-_3)\;,\nonumber\\
&&h_z = {1\over v}(v_1 A_1 + v_2 A_2 + v_3 A_3)\;,\nonumber\\
&& a = (-v_1 v^2_3 A_1 -v_2 v^2_3 A_2 +  v^2_{12}v_3 A_3 -  v^2
v_s A_s)/N_a\;,
\end{eqnarray}
where $v^2 = v^2_1+v^2_2 +v^2_3$ and $N_a^2 = (v_{12}^2v_3^2v^2 +
v^4 v_s^2)$ with $v^2_{12} = v^2_1 +v^2_2$.

We remove $h_w$ and $h_z$ in the Yukawa interaction by making the
following changes of basis,
\begin{eqnarray}\label{mixing}
&&\left (\begin{array}{c}A_1\\A_2\\A_3\\A_s\end{array}\right ) =
\left ( \begin{array}{cccc} v_2/v_{12}&-v_1v_3
v_s/N_A&v_1/v&-v_1v_3^2/N_a\\
-v_1/v_{12}&-v_2v_3
v_s/N_A&v_2/v&-v_2v_3^2/N_a\\
0&v^2_{12}
v_s/N_A&v_3/v&v_{12}^2v_3/N_a\\
0&v^2_{12} v_3/N_A&0&-v^2v_s/N_a\end{array}\right
)\left ( \begin{array}{c}a_1\\a_2\\h_z\\a \end{array}\right )\;,
\nonumber\\
&&\left (\begin{array}{c}h^-_1\\h^-_2\\h^-_3\end{array}\right ) =
\left ( \begin{array}{ccc} v_2/v_{12}&v_1v_3/v v_{12}&v_1/v\\
-v_1/v_{12}&v_2v_3
/v v_{12}&v_2/v\\
0&-v_{12}/v&v_3/v\end{array}\right )\left (
\begin{array}{c}H^-_1\\H^-_2\\h_w \end{array}\right )\;,
\end{eqnarray}
where $N_A^2 = v^2_{12}(v^2_{12}v^2_3 +v_s^2v^2)$. $a_{1,2}$ and
$H^-_{1,2}$ are the physical degrees of freedom for the Higgs
fields. With the same rotation as that for the neutral pseudoscalar,
the neutral scalar Higgs fields $(R_1,R_2,R_3,R_s)^T$ become
$(H_1^0,H_2^0,H^0_3, H^0_4)^T$. Since the invisible axion scale $v_s$
is much larger than the electroweak scale, to a very good
approximation, $N_a = v^2v_s$ and $N_A = v_{12} v v_s$.

In the rotated basis described above, we have the Yukawa
interactions for physical Higgs degrees of freedom as the following
\begin{eqnarray}
L_Y^{(a)} &=& \bar U_L [{\hat M_{u} }\frac{v_1}{v_{12}v_2} - (\hat
M_{u}-V_{CKM}Im(V^\dagger_{CKM})\hat M_u {e^{i\delta}\over
\sin\delta} ) \frac{v_{12}}{v_1v_2} ]U_R(H_1^0+i a_1^0)
\nonumber \\
&+& \bar U_L {\hat M_{u}} U_R [\frac{v_3}{v_{12}v}(H_2^0+i a_2)
-\frac{1}{v}H_3^0+\frac{v_3^2}{v^2 v_s}(H_4^0 + i a )]
\nonumber\\
& -& \bar D_L \hat M_d D_R
[\frac{v_{12}}{v_3v}(H_2^0-ia_2)+\frac{1}{v}H_3^0
+\frac{v_{12}^2}{v^2 v_s}(H_4^0 -i a)  ]
\nonumber\\
&+& \sqrt{2} \bar D_L [V_{CKM}^\dagger \hat
M_{u}\frac{v_1}{v_2v_{12}} -(V_{CKM}^\dagger \hat M_{u}-
Im(V_{CKM}^\dagger)\hat M_u {e^{i\delta}\over \sin\delta})\
\frac{v_{12}}{v_1v_2}]U_R {H^-_1}\nonumber\\
&-&\sqrt{2}\frac{v_3}{v_{12}v} \bar D_L V_{CKM}^\dagger {\hat M_u}
U_R H^-_2 - \sqrt{2}\frac{v_{12}}{vv_3} \bar U_L V_{CKM} \hat M_d
D_R
H_2^+  + h.c.\;,\nonumber\\
 L^{(b)}_Y &=& \bar D_L [{\hat M_{d}
}\frac{v_1}{v_{12}v_2} - (\hat M_d +
V^\dagger_{CKM}Im(V_{CKM})\hat M_d {e^{-i\delta}\over \sin\delta})
\frac{v_{12}}{v_1v_2} ]D_R(H_1^0 - i a_1^0)
\nonumber\\
& +& \bar D_L \hat M_{d} D_R [\frac{v_3}{v_{12}v}(H_2^0-i a_2)
-\frac{1}{v}H_3^0+\frac{v_3^2}{v^2 v_s}(H_4^0 - i a )]
\nonumber\\
&-& \bar U_L {\hat M_u} U_R [\frac{v_{12}}{v_3 v} (H_2^0 + i a_2)
+\frac{1}{v} H_3^0 + \frac{v_{12}^2}{v^2 v_s} (H_4^0 +i a)]
\nonumber\\
& -&\sqrt{2} \bar U_L [V_{CKM} \hat M_{d}\frac{v_1}{v_2v_{12}}
-(V_{CKM} \hat M_d + Im(V_{CKM})\hat M_d {e^{-i\delta}\over \sin\delta})
\frac{v_{12}}{v_1v_{2}}]D_R {H^+_1}\nonumber\\
& +& \sqrt{2}\frac{v_3}{v_{12}v} \bar U_L V_{CKM} {\hat M_d} D_R
H^+_2 + \sqrt{2} \frac{v_{12}}{vv_3} \bar D_LV^\dagger_{CKM}\hat
M_u U_R H_2^-+ h.c.\;.
\end{eqnarray}

Note that the couplings of $a$ and $H^0_4$ to quarks are suppressed by
$1/v_s$, and that only the exchange of $H_1^0$ and $a_1^0$ can induce
tree level FCNC interactions. The FCNC couplings are proportional to
$V_{CKM}Im(V_{CKM}^\dagger)\hat M_u$ and $V^\dagger_{CKM} Im(V_{CKM})\hat M_d$ for
models a) and b), respectively.

We have mentioned before that the identification of the phase
$\delta$ with that in the CKM matrix does not uniquely determine
the full Yukawa coupling. Here we give two often used
parameterizations, the PDG CKM matrix and the original KM matrix
with the CP violating phase indicated by $\delta_{KM}$, to
illustrate the details. In the two cases under consideration, the
phase $\delta$ are identified with $-\delta_{13}$ and
$-\delta_{KM}$, respectively. The differences will show up in the
FCNC of neutral Higgs coupling to quarks which are proportional to
the following quantities,
\begin{eqnarray}
\mbox{PDG}:\;\;&&V_{CKM}Im(V^\dagger_{CKM})\hat M_u  =
-\sin\delta_{13}e^{i\delta_{13}} \left (
\begin{array}{ccc}
c_{13}^2 &  -s_{23}s_{13}c_{13}&-c_{23}s_{13}c_{13}\\
-s_{23}s_{13}c_{13} &  s_{23}^2s_{13}^2 & s_{23}c_{23}s_{13}^2\\
-c_{23}s_{13}c_{13} & s_{23}c_{23}s_{13}^2 & c_{23}^2s_{13}^2
\end{array} \right ) \hat M_u \;,\nonumber\\
&&V_{CKM}^\dagger Im(V_{CKM})\hat M_d =\sin\delta_{13}
e^{-i\delta_{13}}\left (
\begin{array}{ccc}
c_{12}^2 &  s_{12}c_{12}& 0\\
s_{12}c_{12} & s_{12}^2 & 0\\
0 & 0 & 0
\end{array} \right )\hat M_d \;;\nonumber\\
\mbox{KM}:\;\;&&V_{CKM}Im(V^\dagger_{CKM})\hat M_u
=-\sin\delta_{KM}e^{i\delta_{KM}} \left (
\begin{array}{ccc}
0&0&0\\
0&s^2_2&-s_2c_2\\ 0&-s_2c_2&c_2^2
\end{array} \right )\hat M_u \;,\nonumber\\
&&V_{CKM}^\dagger Im(V_{CKM})\hat M_d  =
\sin\delta_{KM}e^{-i\delta_{KM}} \left (
\begin{array}{ccc}
0&0&0\\
0&s^2_3&-s_3c_3\\ 0&-s_3c_3&c_3^2
\end{array} \right )\hat M_d \;.\label{fcnc1}
\end{eqnarray}

\section{Meson and Anti-meson mixing and neutron EDM}

In this section we study some implications for meson and
anti-meson mixing and neutron electric dipole moment.

\subsection{Meson and Anti-meson Mixing}
Meson and anti-meson mixing has been observed previously in
$K^0-\bar K^0$, $B_{d,s}^0-\bar B^0_{d,s}$~\cite{pdg} and in
$D^0-\bar D^0$ very recently~\cite{dmixing}. In the models
considered in the previous section, besides the standard ``box''
diagram contributions to the mixing due to $W$ exchange, there are
also tree level contributions due to the FCNC interactions of
$H^0_1$ and $a_1$.

The interaction Lagrangian for $H_l$ and $a_k$ with quarks
have the following form for both models a) and b),
\begin{eqnarray}\label{Haint}
L = \bar q_i (a^l_{ij} +b^l_{ij}\gamma_5)q_j H_l^0 + i \bar q_i (c^k_{ij} + d^k_{ij}\gamma_5)
q_j a_k\;.
\end{eqnarray}

For the meson and anti-meson mixing, only the FCNC interaction
terms of $H^0_1$ and $a_1$ contribute.  We can write
$a^1=d^1=\alpha$ and $b^1=c^1=\beta$, with $\alpha =
(A+A^\dagger)/2$ and $\beta = (A- A^\dagger)/2$, and $A$ given by:
\begin{eqnarray}
&&\mbox{For a)},\;\; A = V_{CKM}Im(V_{CKM}^\dagger)\hat M_u
{e^{i\delta}\over \sin\delta}{v_{12}\over v_1v_2}\;;\nonumber\\
&&\mbox{For b)},\;\;  A = -V_{CKM}^\dagger Im(V_{CKM})\hat
M_{d}{e^{-i\delta}\over \sin\delta}{v_{12}\over v_1v_2}.
\end{eqnarray}

Using the definition $<0|\bar q_i \gamma^\mu \gamma_5 q_j> = i f_P
p^\mu_P/\sqrt{2m_P}$ and the equation of motion $\bar q_i \gamma_5
q_j = (p_i - p_j)^\mu \bar q_i \gamma_\mu \gamma_5 q_j/(m_i+m_j)$
with $p^P = p_j - p_i$, we obtain the matrix element for $P-\bar
P$ mixing in the vacuum saturation approximation as
\begin{eqnarray}
\nonumber M_{12}&=& \frac{1}{m^2_{H_1}} \left[
(b_{ij}^2-\frac{1}{12}(a_{ij}^2+b_{ij}^2))\frac{f_P^2m_P^3}{(m_i+m_j)^2}
+\frac{1}{12}(b_{ij}^2-a_{ij}^2)f_p^2m_P \right]
\\
&& -\frac{1}{m^2_{a_1}}\left[
(a_{ij}^2-\frac{1}{12}(a_{ij}^2+b_{ij}^2))\frac{f_P^2m_P^3}{(m_i+m_j)^2}
+\frac{1}{12}(a_{ij}^2-b_{ij}^2)f_P^2m_P\right]
\\ \nonumber
&&
+\frac{i2m_{H_1a_1}^2}{m_{H_1}^2m_{a_1}^2}\frac{5a_{ij}b_{ij}}{6}\frac{f_P^2m_P^3}{(m_i+m_j)^2}\;.
\end{eqnarray}
where $m^2_{H_1 a_1}$ parameterizes the mixing between $a_1$ and
$H_1$, that is determined from the Higgs potential $V = m^2_{H_1 a_1}H_1
a_1 + ...$. Since $m^2_{H_1a_1}$ involves new parameters, it can
be made small enough to avoid any conflict with data. We will
neglect their contributions when discussing meson and anti-meson
mixing.  We will come back to this when discussing neutron EDM.

It is obvious from the structure of the flavor changing coupling
in Eq.~(\ref{fcnc1}) that the flavor changing structure for the
PDG and KM cases are different. For PDG case, in model a) there is
tree level contribution from neutral Higgs exchange to $D^0-\bar
D^0$ mixing, but no contribution to $K^0$, $B^0_d$ and $B_s^0$
mixing. In model b), there is only non-zero contribution to
$K^0-\bar K^0$ mixing at the tree level. For the KM case, there is
no tree level contribution to meson mixing in model a). For model
b), there is only non-zero contribution to $B_s^0$ mixing.

In our numerical analysis, we will use the following values for
the relevant parameters. For the CKM matrix elements, we take the
PDG central values with~\cite{pdg}: $s_{12}=0.227$,
$s_{23}=0.042$, $s_{13}=0.004$ and $\sin\delta_{13} =0.84$
(equivalently $s_1=0.227$, $s_2 =0.0358$, $s_3=0.0176$ and
$\sin\delta =0.97$ for the KM parameterization). For the quark
masses, we take~\cite{koide} $m_u \mbox{(1~GeV)}=5 ~\mbox{MeV},
m_d \mbox{(1~GeV)}=10~\mbox{MeV}, m_s \mbox{(1~GeV)}
=187~\mbox{MeV}, m_c(m_c)=1.30~\mbox{GeV},
m_b(m_b)=4.34~\mbox{GeV}, m_t=174~\mbox{GeV}$. For the meson decay
constants, we take~\cite{lattice} $f_K =156~\mbox{MeV}$, $f_D =
201~\mbox{MeV}$, and $f_{B_s} = 260~\mbox{MeV}$.
\\

\noindent{\underline{Models with PDG parameterization}}

We consider the models with PDG parameterization first.

{Model a):} In this case there is mixing only in
$D^{0}-\bar{D^{0}}$ system. Combining the BaBar and Belle
~\cite{dmixing} results the 68\% C.L. range for  $x= {\Delta
m/\Gamma_D}$ is determined to be $(5.5\pm 2.2)\times 10^{-3}$
~\cite{he-dmixing}. Theoretically the elements in $A$ for this
particular case are
$A_{12}=-s_{23}s_{13}c_{13}{{m_cv_{12}}\over{v_1v_2}}$ and
$A_{21}=-s_{23}s_{13}c_{13}{{m_uv_{12}}\over{v_1v_2}}$, which
implies that $a_{12}\sim b_{12}\sim
-s_{23}s_{13}c_{13}{{m_cv_{12}}\over{2v_1v_2}}$. We obtain
\begin{eqnarray}
x&\approx&\frac{5}{12}s_{23}^{2}s_{13}^{2}c_{13}^{2}(
\frac{v_{12}m_{c}}{v_{1}v_{2}})^2\frac{f_{D}^{2}m_{D}}{\Gamma_{D}}
(\frac{m_{D}}{m_{c}+m_{u}})^{2}
(\frac{1}{m^2_{H_{1}}}-\frac{1}{m^2_{a_{1}}})\nonumber\\
&=& 7.5\times10^{-5}\frac{1}{(\sin 2\beta)^2 v_{12}^{2}}
(\frac{1}{m_{H_{1}}^{2}}-\frac{1}{m_{a_{1}}^{2}})(100~\mbox{GeV})^{4}\;.
\end{eqnarray}
where $\tan\beta$ is defined to be ${v_1}/{v_2}$.

It is well known that the SM short distance contribution to the
$D-\bar D$ mixing is small. Long distance contributions can be
much larger, but they suffer from considerable uncertainty. New
physics may contribute significantly~\cite{he-dmixing}. It is
tempting to see if the new contribution in this model can account
for the full measured value. If the effective neutral Higgs mass
$m^2_{eff} = 1/(1/m^2_{H_1} - 1/m^2_{a_1})$ is of order 100 GeV,
one would require $\sin^22\beta v^2_{12} \sim (12)^2$ GeV$^2$.
Since $v_{1,2}$ are related to the top quark mass, with the
assumption that the top quark Yukawa coupling $y_t\leq 1$, one of
them should be large, $\sim$ 240 GeV. Saturating the experimental
central value for $x$, we would have $\sin(2\beta) \sim 0.05$
implying $v_1/v_2$ or $v_2/v_1$ to be of the order of 1/40. If all
vevs are the same order of magnitude, the new contribution does
not produce large enough $x$ to saturate the measured value.
\\

{Model b): } In this case there is mixing only in $K^0- \bar
K^0$ system.  We have
\begin{equation}\label{kkbar}
\frac{\Delta m_{K}}{m_{K}}=4.4\times10^{-12}\frac{1}{\sin^2 2\beta
v_{12}^2}(\frac{1}{m_{H_{1}}^{2}}-\frac{1}{m_{a_{1}}^{2}})(100~\mbox{GeV})^{4}.
\end{equation}
This is to be compared with the experimental data $\Delta
m_{K}/m_{K}=7.0\times10^{-15}$.  It puts strong constraints on
the scalar masses. i.e., the Higgs particles must be at least TeV
scale to suppress the value if $a_1$ and $H_1$ are not degenerate in
mass.
\\

\noindent {\underline{ Models with KM parameterization}}

We now come to models with the original KM parameterization. In
this case, there is no meson and anti-meson mixing in Model a).

{Model b):} There is mixing only in $B_s - \bar B_s$ system. We
have
\begin{equation}
\frac{\Delta
m_{B_{S}}}{m_{B_{s}}}=9.5\times10^{-12}\frac{1}{\sin^22\beta
v_{12}^{2}}(\frac{1}{m_{H_{1}}^{2}}-\frac{1}{m_{a_{1}}^{2}})(100~\mbox{GeV})^{4}.
\end{equation}

Experimental value $\Delta m_{B_s}=17.4~\mbox{ps}^{-1}$ implies
$\Delta m_{B_s}/m_{B_s}=2.1\times10^{-12}$. It has been shown in
Ref.~\cite{lenz06} that the New Physics contribution to
$\Delta m_{B_s}$ can be up to 10\%. To obtain the lowest Higgs
boson mass, we maximize $\sin2\beta=1$ which requires $v_1 = v_2$.
Taking $v_{1,2,3}$ to be all equal, the Higgs boson mass can be as
low as $300$ GeV. With smaller $v_{1,2}$ or non-equal $v_1$ and
$v_2$ would lead to more stringent bound on Higgs mass.

\subsection{The neutron EDM}

The neutron EDM can also provide much information on the model
parameters. The standard model predicts a very small~\cite{hmp}
$d_n$ ($<10^{-31}e$ cm). The present experimental upper bound on
neutron EDM $d_n$ is very tight ~\cite{pdg}: $|d_n|<0.63\times
10^{-25} e$ cm. We now study whether neutron EDM can reach its
present bound after imposing the constraints from meson and
anti-meson mixing discussed in the previous section.

In the models we are studying, the quark EDMs will be generated at
loop levels due to mixing between $a_i$ and $H_i$. From Higgs
potential given earlier, we find the mixing parameters,
\begin{eqnarray}
m_{H_{1}a_{1}}^{2}&=&[(\lambda_{6}-\lambda_{7})v_{1}v_{2}-\lambda_{5}(v_{1}^{2}-v_{2}^{2})
\cos\delta]\sin\delta\;,
\nonumber\\
m_{H_{1}a_{2}}^{2}&\simeq&
-\frac{f_{13}\sin(\delta+\delta_{s})vv_{s}}{\sqrt{2}v_{2}}\;,
\nonumber\\
m_{H_{2}a_{1}}^{2}&\simeq&
\frac{1}{2v_{2}v}[-2\lambda_{5}v_{1}v_{3}v_{2}^{2}\sin2\delta
+2(-\lambda_{6}v_{1}^{2}-\lambda_{7}v_{2}^{2}+(\lambda_{8}+d_{12})
v_{12}^{2})v_{2}v_{3}\sin\delta
\nonumber\\
&&
+\sqrt{2}f_{13}v^2v_{s}\sin(\delta+\delta_s)]\;,
\nonumber\\
m_{H_{3}a_{1}}^{2}&=&\frac{v_{12}}{v}[2\lambda_{5}v_{1}v_{2}\cos(\theta_{1}-\theta_{2})
+\lambda_{6}v_{1}^{2}+\lambda_{7}v_{2}^{2}+\lambda_{8}v_{3}^{2}]\sin\delta.
 \end{eqnarray}
Note that all the parameters above are zero if $\sin\delta = 0$.

The one loop contributions to the neutron EDM are suppressed for the
usual reason of being proportional to light quarks masses to the third
power for diagram in which the internal quark is the same as the
external quark.  In model a) with PDG parameterization, there is a
potentially large contribution when there is a top quark in the
loop. However, the couplings to top are proportional to $s_{13}$,
therefore the contribution to neutron EDM is much smaller than the
present upper bound. We will not discuss them further.

It is well known that exchange of Higgs at the two loop level may
be more important than the one loop contribution, through the
quark EDM $O_q^\gamma$~\cite{bz}, quark color EDM $O^C_q$~\cite{bz},
and the gluon color EDM $O^C_g$~\cite{weinberg2} defined as
\begin{eqnarray}
O^\gamma_q=-\frac{d_q}{2}i\bar{q}\sigma_{\mu\nu}\gamma_5F^{\mu\nu}q\;,
\;\;O_q^C=-\frac{f_q}{2}ig_s\bar{q}\sigma_{\mu\nu}\gamma_5G^{\mu\nu}q\;,
\;\;O_g^C=-\frac{1}{6}Cf_{abc}G_{\mu\nu}^aG_{\mu\alpha}^b\tilde{G}^c_{\nu\alpha}\;,
\end{eqnarray}
where $F^{\mu\nu}$ is the photon field strength, $G^{\mu\nu}$ is the gluon field strength and
$\tilde{G}^{\mu\nu}=\frac{1}{2}\epsilon_{\mu\nu\alpha\beta}G^{\alpha\beta}$.

In the valence quark model, the quark EDM and color EDM
contributions to the neutron EDM $d_n$ are given by~\cite{hmp}
\begin{eqnarray}
d_n^\gamma = \eta_d \left[ \frac{4}{3}d_d-\frac{1}{3}d_u
\right]_\Lambda\;, \;\; d_n^C = e \eta_f \left[
\frac{4}{9}f_d+\frac{2}{9}f_u \right]_\Lambda\;,
\end{eqnarray}
where~\cite{darwin} $\eta_d =\left(
\frac{\alpha_s(M_Z)}{\alpha_s(m_b)} \right)^{16/23} \left(
\frac{\alpha_s(m_b)}{\alpha_s(m_c)} \right)^{16/25} \left(
\frac{\alpha_s(m_c)}{\alpha_s(\Lambda)} \right)^{16/27}\approx
0.166$ and $\eta_f =\left( \frac{\alpha_s(M_Z)}{\alpha_s(m_b)}
\right)^{14/23} \left( \frac{\alpha_s(m_b)}{\alpha_s(m_c)}
\right)^{14/25} \left( \frac{\alpha_s(m_c)}{\alpha_s(\Lambda)}
\right)^{14/27}\frac{\alpha_s(M_Z)}{\alpha_s(\Lambda)}\approx
0.0117$ are the QCD running factors from scale $m_Z$ to the hadron
scale $\Lambda$.

A naive dimensional analysis (NDA) estimate gives the gluon color
EDM contribution to  the neutron EDM as the following
\begin{equation}
d_n\approx \frac{eM}{4\pi}\xi C,
\end{equation}
where $M=4\pi f_\pi=1190~\mbox{MeV}$ is the scale of chiral
symmetry breaking. The QCD running factor is~\cite{tc}
$\xi=\left(\frac{g(\Lambda)}{4\pi}\right)^3
\left(\frac{\alpha_s(m_b)}{\alpha_s(m_t)}\right)^{-54/23} \left(
\frac{\alpha_s(m_c)}{\alpha_s(m_b)}\right)^{-54/25}
\left(\frac{\alpha_s(\Lambda)}{\alpha_s(m_c)}\right)^{-54/27}\approx
1.2\times 10^{-4}$.

The two loop contribution to $d_q$,  $f_q$  and $C$ are given by
\begin{eqnarray}
d_q&=&\frac{e\alpha_{em}Q_q}{24\pi^3}m_q G(q)\;,\;\; f_q =
\frac{\alpha_s}{64 \pi^3}m_q G(q)\;,\;\;C=\frac{1}{8\pi}H(g) \;,
\end{eqnarray}
where $Q_q$ is the charge of $q$ quark and
\begin{eqnarray}\label{fgh}
&&\nonumber
G(q)=\left[(f(\frac{m^2_t}{m^2_{H_l}}) -
f(\frac{m^2_t}{m^2_{a_k}}))Im
Z_{tq}^{lk}+(g(\frac{m^2_t}{m^2_{H_l}})-g(\frac{m^2_t}{m^2_{a_k}}))
Im Z_{qt}^{lk}\right]\;,
\\
&&
H(g) = (h(\frac{m^2_t}{m^2_{H_l}}) -
h(\frac{m^2_t}{m^2_{a_k}}))ImZ_{tt}^{lk}\;,
\end{eqnarray}
where $Im Z_{ij}$ is defined through $Im Z_{ij}^{lk}=2{a^l_{ii}d^k_{jj}\lambda_{lk}}/(m_im_j)$
with $a^l, d^k$ defined by Eq.~(\ref{Haint})
and $\lambda_{lk}=m^2_{H_l a_k}/(m^2_{H_l} - m^2_{a_k})$ is a mixing factor
depending on the neutral Higgs bosons exchanged in the loop.

The functions $f(z)$, $g(z)$ and
$h(z)$ are given by
\begin{eqnarray}
\nonumber
f(z)&=&\frac{z}{2}\int^1_0dx\frac{1-2x(1-x)}{x(1-x)-z}\ln\frac{x(1-x)}{z}\;,
\\
g(z)&=&\frac{z}{2}\int^1_0dx\frac{1}{x(1-x)-z}\ln\frac{x(1-x)}{z}\;,\\
h(z)&=&\frac{z^2}{2}\int^1_0 dx\int^1_0 du
\frac{u^3x^3(1-x)}{\left[ z x(1-ux)+(1-u)(1-x) \right]^2}\;.
\end{eqnarray}
Numerically we find that functions $(f, g, h)$ change slowly
from $(0.5,1,0.1)$ to $(0.2,0.2,0.03)$ when Higgs masses are
increased from 100 GeV to 1 TeV.
\\

\noindent {\underline{Models with PDG parameterization}}

{Model a):} The 2-loop contributions to the neutron EDM due to
the Higgs bosons exchange in the loop are proportional to the
mixing factor $\lambda_{lk}(f,\;g,\;h)$ . We take these factors to
be approximately equal to estimate the contributions from different
Higgs exchange.

If using the parameters which produce $D$ mixing, i.e.,
$\tan\beta=40, v_{12}\sim 240~\mbox{GeV}$ and $v_3\sim
10~\mbox{GeV}$ and Higgs around $100$ GeV are used, we find that
the dominant contribution is from $H_3, a_1$ exchange,
\begin{eqnarray}
d_n\approx -1.5 \times 10^{-25}\frac{m^2_{H_3a_1}}{m_{H_3}^2-m_{a_1}^2}\mbox{e cm}\;.
\end{eqnarray}

If all vevs are of the same order, i.e., taking $v_1=v_2=v_3$ with
Higgs mass of order $100$ GeV, we have
\begin{eqnarray}
d_n \approx 8 \times
10^{-26}\frac{m^2_{H_3a_1}}{m_{H_3}^2-m_{a_1}^2}\mbox{e cm}\;.
\end{eqnarray}

{Model b):} In this case $H_1, a_1$ do not couple to $\bar tt$,
so the two loop contribution to quark EDM and quark and gluon
color EDM from the $H_1, a_1$ are small.

The contributions to neutron EDM are about the same from the
$H_1, a_{2}$ and $H_{2,3}, a_1$ exchange, with different mixing
factors. Explicitly as an example, for the case $H_1, a_2$
exchange with the Higgs mass taken to be 1 TeV, as high as
allowed by  $K^0-\bar K^0$ mixing, we have
\begin{equation}
d_n\approx -1 \times 10^{-26}\frac{m^2_{H_1a_2}}{m_{H_1}^2-m_{a_2}^2}\mbox{e cm}\;.
\end{equation}
If $m^2_{H_1 a_2}$ is not too much smaller than $m^2_{H_1,a_2}$,
the neutron EDM can be close to the upper bound.
\\

\noindent {\underline{Models with KM parameterization}}

{ Model a):} In this case there are no constraints from meson
mixing, the Higgs mass can be low. If all vevs are of the same
order, i.e. taking $v_1=v_2=v_3$ with Higgs mass of order 100 GeV,
we have the main contribution come from $H_1, a_2$ exchange,
\begin{eqnarray}
d_n \approx 5 \times 10^{-26}\frac{m^2_{H_1a_2}}{m_{H_1}^2-m_{a_2}^2}\mbox{e cm}\;.
\end{eqnarray}

{Model b):} Similar to the case for Model b) as in the PDG
parameterization case, the contributions from the $H_1, a_1$ exchange
 are small. Taking the vevs to be same order and Higgs
mass to be of the order of 100 GeV, we find the contributions from $H_1,
a_2$ exchange and $H_{2,3}, a_1$ exchange are comparable. For the
case $H_1, a_2$ exchange, the contribution is given by
\begin{equation}
d_n\approx 5 \times
10^{-26}\frac{m^2_{H_1a_2}}{m_{H_1}^2-m_{a_2}^2}\mbox{e cm}\;.
\end{equation}
If one takes the Higgs mass to be 300 GeV as that from $B_s-\bar
B_s$ mixing, the neutron EDM will be smaller.

\section{Discussions and Conclusions}

In our previous discussions, we have not considered Yukawa
coupling for the lepton sector. An analogous study can be carried
out. If one introduces right handed neutrinos, see-saw mechanism
can be applied to generate small neutrino masses if the axion
scale $v_s$ is identified with the see-saw scale. We briefly
discuss two classes of models parallel to the quark sector before
conclusion.

{Model a):} The PQ charges for lepton doublet $L_L$, electron
$e_R$ and neutrino $\nu_R$ are assigned as: $L_L ( 0)$, $e_R (-1)$
and $\nu_R (-1)$. The Yukawa couplings are then
\begin{eqnarray}
L = \bar L_L (Y_1 H_1 + Y_2 H_2 e^{i \delta}) \nu_R + \bar L_L Y_3
\tilde H_3 e_R + \bar \nu^C_R Y_s S e^{i(\delta +\delta_s)} \nu_R
+ h.c.
\end{eqnarray}
In this case the mass matrices in  $L_m = - \bar e_L M_e e_R -
\bar \nu_L M_D \nu_R - (1/2) \bar \nu^C_R M_R \nu_R$ can be
written as
\begin{eqnarray}
M_l = - {1\over \sqrt{2}} Y_3 v_3,\;\; M_D = -{1\over
\sqrt{2}}(Y_1 v_1 + Y_2 v_2 e^{i\delta} ),\;\;M_R = - \sqrt{2} Y_s
v_s e^{i(\delta +\delta_s)}.
\end{eqnarray}

The charged current mixing matrix in the lepton sector, the
Pontecove-Maki-Nakagawa-Sakata (PMNS) matrix~\cite{pmns},
$V_{PMNS}$ similar to the $V_{CKM}$ matrix is given by $V_{PMNS} =
V^e_L V^{\nu\dagger}_L$, where $V_L^e$ and $V^\nu_L$ are defined
by: $M_e =V^{e\dagger}_L  \hat M_e V^e_R$ and $M_\nu = -M_D
M^{-1}_R M^T_D = V^{\nu\dagger}_L \hat M_\nu V^{\nu*}_L$ with
$\hat M_e$ and $\hat M_\nu$ the charge lepton and light neutrino
eigen-mass matrices.

{Model b):} The PQ charges for lepton doublet $L_L$, electron
$e_R$ and neutrino $\nu_R$ are assigned as: $L_L ( 0)$, $e_R (+1)$
and $\nu_R (+ 1)$. The Yukawa couplings are
\begin{eqnarray}
L = \bar L_L Y_3 H_3 \nu_R + \bar L_L (Y_1 \tilde H_1 + Y_2 \tilde
H_2 e^{-i\delta}) e_R + \bar \nu^C_R Y_s S^\dagger e^{-i(\delta
+\delta_s)} \nu_R + h.c.\;,
\end{eqnarray}
and
\begin{eqnarray}
M_l = - {1\over \sqrt{2}} (Y_1 v_1 + Y_2 v_2 e^{-i\delta}),\;\;
M_D = -{1\over \sqrt{2}} Y_3 v_3,\;\;M_R = -\sqrt{2} Y_s v_s
e^{-i(\delta +\delta_s)}\;.
\end{eqnarray}

From the above we see that, in general there are CP violation in
the mixing matrix $V_{PMNS}$, and the source is the same as that
in the Higgs potential. But the identification of the phase
$\delta$ with the phase in the $V_{PMNS}$ becomes more complicated
due to the appearance of $M_R$. The related details will be
discussed elsewhere.

We have proposed that the CP violating phase in the CKM mixing
matrix to be the same as that causing spontaneous CP violation in
the Higgs potential.  Specific multi-Higgs doublet models have
been constructed to realize this idea. There are flavor changing
neutral current mediated by neutral Higgs bosons at the tree
level. However, even when the Higgs boson masses are set to be
very large, the phase in the CKM matrix can be made finite and CP
violating effects will not disappear unlike in other models of
spontaneous CP violation where the CP violation disappear when
Higgs boson masses become large. Another interesting feature of
this model is that the FCNC Yukawa couplings are fixed in terms of
the quark masses and CKM mixing angles, making phenomenological
analysis much easier.

We have studied some implications for meson-anti-meson mixing,
including recent data on $D-\bar D$ mixing, and neutron electric
dipole moment. We find that the neutral Higgs
boson masses can be at the order of 100 GeV. The neutron
EDM can be close to the present experimental upper bound.

\noindent {\bf Acknowledgments}$\,$ This work was supported in
part by the National Science Council and the National Center for
Theoretical Sciences, and by the U.S. Department of Energy under
Grants No DE-FG02-96ER40969. We thank Jon Parry for pointing out a
typo in our first version on the arXiv.


\begin{thebibliography}{01}


\bibitem{pdg} W-M Yao {\it et al} 2006 J. Phys. G: Nucl. Part. Phys. {\bf 33} 1.

\bibitem{cab} N. Cabbibo, Phys. Rev. Lett. {\bf 10}, 531(1963).

\bibitem{km} 
  M.~Kobayashi and T.~Maskawa,
  Prog.\ Theor.\ Phys.\  {\bf 49}, 652 (1973).


\bibitem{tdlee}
  T.~D.~Lee,
  Phys.\ Rev.\  D {\bf 8}, 1226 (1973);
  T.~D.~Lee,
  Phys.\ Rept.\  {\bf 9}, 143 (1974).

\bibitem{weinberg1}
  S.~Weinberg,
  Phys.\ Rev.\ Lett.\  {\bf 37}, 657 (1976);
  G.~C.~Branco,
  Phys.\ Rev.\ Lett.\  {\bf 44}, 504 (1980).



\bibitem{problem}
  D.~Chang, X.~G.~He and B.~H.~J.~McKellar,
  Phys.\ Rev.\  D {\bf 63}, 096005 (2001)
  [arXiv:hep-ph/9909357];
  G.~Beall and N.~G.~Deshpande,
  Phys.\ Lett.\  B {\bf 132}, 427 (1983);
  I.~I.~Y.~Bigi and A.~I.~Sanda,
  Phys.\ Rev.\ Lett.\  {\bf 58}, 1604 (1987).

\bibitem{bigi}
  R.~Akhoury and I.~I.~Y.~Bigi,
  Nucl.\ Phys.\  B {\bf 234}, 459 (1984).

\bibitem{Ball:1999mb}
  P.~Ball, J.~M.~Frere and J.~Matias,
  Nucl.\ Phys.\  B {\bf 572}, 3 (2000)
  [arXiv:hep-ph/9910211].


\bibitem{other}
  G.~C.~Branco and R.~N.~Mohapatra,
  Phys.\ Lett.\  B {\bf 643}, 115 (2006)
  [arXiv:hep-ph/0607271];
  G.~C.~Branco, D.~Emmanuel-Costa and J.~C.~Romao,
  Phys.\ Lett.\  B {\bf 639}, 661 (2006)
  [arXiv:hep-ph/0604110].



\bibitem{pq} 
  R.~D.~Peccei and H.~R.~Quinn,
  Phys.\ Rev.\  D {\bf 16}, 1791 (1977):
  R.~D.~Peccei and H.~R.~Quinn,
  Phys.\ Rev.\ Lett.\  {\bf 38}, 1440 (1977).


\bibitem{hvgn}
  X.~G.~He and R.~R.~Volkas,
  Phys.\ Lett.\  B {\bf 208}, 261 (1988)
  [Erratum-ibid.\  B {\bf 218}, 508 (1989)];
  C.~Q.~Geng, X.~D.~Jiang and J.~N.~Ng,
  Phys.\ Rev.\  D {\bf 38}, 1628 (1988).


\bibitem{invisible}
A.R. Zhitnitsky, Sov. J. Nucl. Phys. {\bf 31}, 260(1980);
  M.~Dine, W.~Fischler and M.~Srednicki,
  Phys.\ Lett.\  B {\bf 104}, 199 (1981).

\bibitem{kk} J. E. Kim, Phys. Rev. Lett. {\bf 43} (1979) 103;
M. Shifman, A. Vainshtein, V. Zakharov, Nucl. Phys. {\bf B166}
(1980) 493.


\bibitem{dmixing}
B. Aubert {\it et al.} [BaBar Collaboration],  arXiv:
hep-ex/0703020, BABAR-PUB-07/019, SLAC-PUB-12385, SCIPP-07/01; M.
Staric, Talk presented at XLII Rencontres de Moriond, La Thuile,
Italy, 10-17 March, 2007; K. Abe {\it et al.} [Belle
Collaboration], arXiv: hep-ex/0703036.

\bibitem{he-dmixing} M. Chiuchini, {\it et al} arXiv:hep-ph/0703204;
  X.~G.~He and G.~Valencia,
  arXiv:hep-ph/0703270;
  C.~H.~Chen, C.~Q.~Geng and T.~C.~Yuan,
  arXiv:0704.0601 [hep-ph];
  Z.~z.~Xing and S.~Zhou,
  arXiv:0704.0971 [hep-ph];
  P.~Ball,
  arXiv:0704.0786 [hep-ph]; K. Babu et al.,
  Phys.\ Lett.\  B {\bf 205}, 540 (1988).


\bibitem{koide}
  H.~Fusaoka and Y.~Koide,
  Phys.\ Rev.\  D {\bf 57}, 3986 (1998)
  [arXiv:hep-ph/9712201].


\bibitem{lattice} M. Okamoto,
PoS(LAT2005)013[arXiv:hep-lat/0510113].

\bibitem{lenz06}
A. Lenz and U. Nierste, arXiv: hep-ph/0612167;
X.~G.~He and G.~Valencia,
  Phys.\ Rev.\  D {\bf 74}, 013011 (2006)
  [arXiv:hep-ph/0605202];
 K.~Cheung, C.~W.~Chiang, N.~G.~Deshpande and J.~Jiang,
  arXiv:hep-ph/0604223.



\bibitem{hmp}
  N.~G.~Deshpande, G.~Eilam and W.~L.~Spence,
  Phys.\ Lett.\  B {\bf 108}, 42 (1982);
  X.~G.~He, B.~H.~J.~McKellar and S.~Pakvasa,
  Int.\ J.\ Mod.\ Phys.\  A {\bf 4}, 5011 (1989)
  [Erratum-ibid.\  A {\bf 6}, 1063 (1991)];
  B.~H.~J.~McKellar, S.~R.~Choudhury, X.~G.~He and S.~Pakvasa,
  Phys.\ Lett.\  B {\bf 197}, 556 (1987).



\bibitem{bz} S.~M.~Barr and A.~Zee,
  Phys.\ Rev.\ Lett.\  {\bf 65}, 21 (1990)
  [Erratum-ibid.\  {\bf 65}, 2920 (1990)];
J.F. Gunion and D. Wyler, Phys. Lett. B {\bf 248}, 170 (1990).


\bibitem{weinberg2} S.~Weinberg, Phys. Rev. Lett. {\bf 63}, 2333 (1989); Phys. Rev. D{\bf 42}, 860
(1990).

\bibitem{darwin} D. Chang, W.-Y. Keung, and T. C. Yuan, Phys. Lett. B {\bf 251},
608 (1990); D.~Chang, et al., Phys. Rev. D {\bf 46}, 3876 (1992).

\bibitem{tc} E.~Braaten, C.-S. Li, and T.-C. Yuan, Phys. Rev. Lett. {\bf 64}, 1709 (1990); Phys. Rev.
D {\bf 42}, 276 (1990).

\bibitem{pmns} B. Pontecovo, Sov. Phys. JETP 6, 429(1957); Z.
Maki, M. Nakagawa and S. Sakata, Prog. Theor. 28, 870(1962).


\end{thebibliography}
\end{document}